\documentclass[conference]{IEEEtran}
\IEEEoverridecommandlockouts

\usepackage[english]{babel}
\usepackage{amsthm}
\usepackage{titlesec}
\usepackage[ruled,vlined]{algorithm2e}
\usepackage{algcompatible}
\usepackage{graphicx}
\usepackage{float}
\usepackage{caption}
\usepackage{tikz}
\usepackage{changepage}
\usepackage[utf8]{inputenc}
\usepackage{pgfplots} 
\usepackage{pgfgantt}
\usepackage{pdflscape}
\usepackage[export]{adjustbox}
\pgfplotsset{compat=newest} 
\pgfplotsset{plot coordinates/math parser=false}
\pgfplotsset{compat=1.18}
\usepackage[justification=centering]{caption}
\newtheorem{innercustomgeneric}{\customgenericname}
\providecommand{\customgenericname}{}
\newcommand{\newcustomtheorem}[2]{%
  \newenvironment{#1}[1]
  {%
   \renewcommand\customgenericname{#2}%
   \renewcommand\theinnercustomgeneric{##1}%
   \innercustomgeneric
  }
  {\endinnercustomgeneric}
}

\newcustomtheorem{customthm}{Theorem}
\newcustomtheorem{customlemma}{Lemma}

\usepackage{amsmath}
\usepackage{acronym}
\usepackage{amssymb}
\usepackage{mathtools}
\usepackage{url}
\usepackage{graphicx}  
\usepackage{float}  

\newcommand{\xx}{\mathbf{x}}

\newcommand{\pp}{\mathbf{p}}

\newcommand{\dd}{\mathbf{d}}
\DeclareMathOperator*{\minimize}{\text{minimize}}
\DeclareMathOperator*{\maximize}{\text{maximize}}

\newcommand{\HH}{\mathbf{H}}

\newcommand{\hh}{\mathbf{h}}

\newcommand{\XX}{\mathbf{X}}
\newcommand{\RR}{\mathbf{R}}

\newcommand{\Tr}{\text{Tr}}
\newcommand{\diag}{\text{diag}}

\newcommand{\hermit}{\mathsf{H}}

\usepackage{accents}

\newcommand{\PPsi}{\boldsymbol{\Psi}}
\newcommand{\ppsi}{\boldsymbol{\psi}}

\acrodef{SISO}[SISO]{single-input single-output}
\acrodef{AP}[AP]{access point}
\acrodef{UE}[UE]{user equipment}
\acrodef{ULA}[ULA]{uniform linear array}
\acrodef{CPU}[CPU]{central processing unit}
\acrodef{LoS}[LoS]{line of sight}
\acrodef{NLoS}[NLoS]{non line of sight}
\acrodef{RCS}[RCS]{radar cross section}
\acrodef{AoD}[AoD]{angle of departure}
\acrodef{AoA}[AoA]{angle of arrival}
\acrodef{CRB}[CRB]{Cramer-Rao bound}
\acrodef{FIM}[FIM]{Fisher information matrix}
\acrodef{AN}[AN]{artificial noise}
\acrodef{SINR}[SINR]{signal-to-interference-plus-noise ratio}
\acrodef{SNR}[SNR]{signal-to-noise ratio}
\acrodef{QoS}[QoS]{quality of service}
\acrodef{SDR}[SDR]{semi-definite relaxation}
\acrodef{SDP}[SDP]{semi-definite relaxation}
\acrodef{ISAC}[ISAC]{integrated sensing and communications}
\acrodef{PLS}[PLS]{physical layer security}
\acrodef{CSI}[CSI]{channel state information}
\acrodef{MUI}[MUI]{multi-user interference}
\acrodef{RIS}[RIS]{Reconfigurable intelligent surface}
\acrodef{SIMO}[SIMO]{Single Input Multiple Output}
\acrodef{BS}[BS]{base station}
\acrodef{CEE}[CEE]{channel estimation error}
\acrodef{CCP}[CCP]{convex concave procedure}
\acrodef{MRT}[MRT]{maximum-ratio transmission}
\begin{document}

\title{Malicious Reconfigurable Intelligent Surfaces: \\ How Impactful can Destructive Beamforming be?}


        
\author{Steven~Rivetti,~\IEEEmembership{Student Member,~IEEE},
        Özlem~Tu$\Breve{\text{g}}$fe~Demir,~\IEEEmembership{Member,~IEEE}\\
        Emil~Björnson,~\IEEEmembership{Fellow,~IEEE}, 
        Mikael~Skoglund,~\IEEEmembership{Fellow,~IEEE}
       \thanks{
This work was supported by the SUCCESS project (FUS21-0026), funded by the Swedish Foundation for Strategic Research.}
        \thanks{Steven Rivetti, Emil Björnson, and Mikael Skoglund   are with the School of Electrical Engineering and Computer Science (EECS), KTH Royal Institute of Technology, 11428 Stockholm, Sweden. Özlem Tu$\Breve{\text{g}}$fe Demir is with the Department of Electrical-Electronics Engineering, TOBB University of Economics and Technology, Ankara, Türkiye. } \vspace{-2mm}}

\maketitle

\begin{abstract}
Reconfigurable intelligent surfaces (RISs) have demonstrated significant potential for enhancing communication system performance if properly configured. However, a RIS might also pose a risk to the network security. In this letter, we explore the impact of a malicious RIS on a multi-user multiple-input single-output (MISO) system when the system is unaware of the RIS's malicious intentions. The objective of the malicious RIS is to degrade the \ac{SNR} of a specific \ac{UE}, with the option of preserving the SNR of the other UEs, making the attack harder to detect. To achieve this goal, we derive the optimal RIS phase-shift pattern, assuming perfect channel state information (CSI) at the hacker. We then relax this assumption by introducing CSI uncertainties and subsequently determine the RIS's phase-shift pattern using a robust optimization approach. Our simulations reveal a direct proportionality  between the performance degradation caused by the malicious RIS and the number of reflective elements, along with resilience toward CSI uncertainties.

\end{abstract}


\section{Introduction}
\acp{RIS} provide intelligent control over radio propagation environments through the utilization of nearly passive integrated electronic circuits that manipulate incoming waves. Based on its significant potential to improve coverage, spectral and energy efficiency, as well as localization performance in cellular networks, \acp{RIS} are acknowledged as a pivotal technology for sixth-generation (6G) networks \cite{alexandropoulos2021reconfigurable}. However, the envisioned low-cost and easy deployment of RISs introduces the risk that these surfaces are exploited as malicious attackers that might degrade the communication performance of user equipments (UEs) \cite{wang2022wireless}. This can be done without generating interfering signals, as in traditional jamming, which might make the attacks hard to detect.
While much of the research on \ac{RIS} has focused on its positive contributions, the massive and cost-effective deployment of \acp{RIS} is anticipated to give rise to security issues when RIS functions as an untrusted component \cite{lin2023pain}. It is critically important to comprehend and analyze all the possible ways in which an \ac{RIS} can be used to exploit the system vulnerabilities and what damages it can cause \cite{hu2023exploiting}.
The consequence of this security threat posed by \acp{RIS} is an increasing interest in the development of techniques aimed at counteracting the action of a malicious \ac{RIS} \cite{huang2023disco}.
The common practice in wireless security analysis is to consider the worst-case scenario where the hacker has perfect \ac{CSI} on all the channels.
While this approach establishes the theoretical limits, it is of practical import to look into the effect that \acp{CEE} have on the \ac{RIS}'s potential malicious actions.

\acp{RIS} are usually envisioned to be deployed as an integral component of the network, enabling the deployment of defensive mechanisms against malicious entities, such as eavesdroppers or jammers.
A comprehensive examination of strategies that enhance the \ac{PLS} can be found in \cite{khalid2023reconfigurable}: the authors present different \ac{RIS}-based design solutions with which the \ac{PLS} of a 6G network can be increased. In \cite{alexandropoulos2023counteracting}, the authors consider a scenario with two \acp{RIS} whereof only one is legitimate, as well as the presence of an eavesdropper amongst the receiving UEs.
Their main objective is to maximize the secrecy rate by incorporating artificial noise into the transmitted waveform.
On the other hand, in \cite{wang2022wireless}, they analyze the signal leakage towards a malicious \ac{RIS}.
Contrary to the usual approach, \cite{lin2023pain} optimizes the attack of a malicious active \ac{RIS}, aimed  at degrading the signal-to-noise ratio (SNR) of a single receiving device.
Assuming perfect \ac{CSI}, the \ac{RIS} and \ac{BS} beamforming vectors are obtained, both in the presence and absence of cooperation between these two parties.

\subsection{Contributions}
Contrary to the existing literature, where a \ac{RIS} is used as a performance booster or as a tool to protect the network from eventual jammers, our work highlights how a \ac{RIS} might be used as a silent attacker that causes harm without transmitting any signals---a feature that makes it hard to detect.
We investigate the impact that the number of \ac{RIS} reflective elements has on the malicious \ac{RIS}'s silent jamming action and how resilient  this action is towards \acp{CEE}.
We first assume perfect \ac{CSI} at the hacker and derive the \ac{RIS} phase-shift pattern that degrades the \ac{SNR} of a designated UE, with the option of introducing minimum \ac{SNR} constraints for the other UEs, making the silent attack even harder to detect as the channel quality of every UE, but one is preserved.
We then introduce \acp{CEE} onto the static path, as the \ac{RIS} never interacts with this channel.
Adopting a bounded error model, we then recast the previously defined optimization problems into robust ones. To the best of the authors' knowledge, this is the first work characterizing the impact of \acp{CEE} onto a malicious \ac{RIS}'s \ac{SNR}-minimizing action.
We provide numerical results highlighting how the impact of the attack depends on the number of RIS elements.
The impact of \acp{CEE} is then assessed by comparing the robust optimization solutions to their perfect-\ac{CSI} counterparts.

\emph{Notation}: Boldface lowercase and uppercase letters denote vectors and matrices, with the symbols $(\cdot)^\top$ and $(\cdot)^\hermit$ representing the transpose and Hermitian transpose operators, respectively.
The trace of the matrix $\XX$ is denoted by $\Tr(\XX)$, while $\diag(\xx)$ represents the stacking of a vector $\xx$ onto the main diagonal of a matrix.
The symbols $|\cdot|$ and $||\cdot||$ denote the absolute value and Euclidean norm, respectively.
The space of $M \times N$ complex matrices is denoted by $\mathbb{C}^{M \times N}$ and
$h(n)$ represents the $n$-th element of the vector $\hh$.

\begin{figure}[t]
\begin{center}		  \resizebox{0.5\textwidth}{!}{
    \input{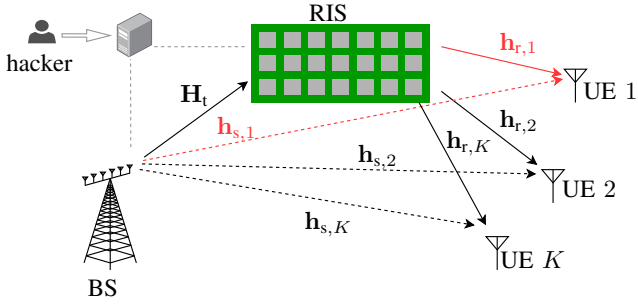}
  }
\vspace{-6mm}
			\caption{A RIS-aided communication system where a hacker has hacked into the RIS's control center.} \label{simple example}
		\end{center}
  \vspace{-4mm}

\end{figure}

\section{System Model}\label{system model}
We consider the \ac{RIS}-assisted system shown in Fig.~\ref{simple example}, where a \ac{BS} with $M$ antennas serves $K$ single-antenna \acp{UE}.
 Unbeknownst to the transmitter, the \ac{RIS} controller has been hacked with the goal of degrading the \ac{SNR} of  one \ac{UE}, here assumed to be \ac{UE} $1$ without loss of generality. To keep its malicious intentions undetected, the RIS should simultaneously guarantee all other \acp{UE} a minimum \ac{SNR}. The \ac{RIS} is equipped with $N$ passive reflective elements, whose passive beamforming action is described by  $\ppsi = [\psi_1 , \ldots ,\psi_N]^\top \in \mathbb{C}^N$, where $|\psi_n|=1$, for $n=1,\ldots,N$. 
The static path between the \ac{BS} and \ac{UE} $k$ is denoted by $\mathbf{h}_{\text{s},k}\in \mathbb{C}^{M}$. The channel between the \ac{BS} and \ac{RIS} is described by  $\HH_{\text{t}}\in \mathbb{C}^{M \times N}$, while the channel between the \ac{RIS} and \ac{UE} $k$ is $\hh_{\text{r},k}\in \mathbb{C}^N$.
We adopt the cascaded channel model to define the overall channel between the \ac{BS} and \ac{UE} $k$ as
\begin{align}\label{cascaded channel}
\hh_{k}=\hh_{\text{s},k}+\HH_\text{t}\HH_{\text{r},k}\ppsi,  
\end{align}
where $\HH_{\text{r},k}=\diag(\hh_{\text{r},k}) \in \mathbb{C}^{N \times N}$ is a diagonal matrix. The \ac{BS} applies a fixed precoding vector $\pp_k \in \mathbb{C}^{M}$ to UE~$k$'s data symbol. The effective end-to-end \ac{SISO} channel to UE~$k$ then becomes
\begin{align}\label{cascaded channel2}
    h_k = \pp^{\top}_k\hh_k = \underbrace{\pp^{\top}_k\hh_{\text{s},k}}_{\triangleq 
h_{\text{s},k}}+\underbrace{\pp^{\top}_k\HH_{\text{t}}\HH_{\text{r},k}}_{\triangleq\Breve{\hh}_k^{\hermit}}\ppsi = h_{\text{s},k} + \Breve{\hh}_k^{\hermit}\ppsi.
\end{align}
The malicious RIS wants to compute its passive beamforming vector $\ppsi$ with the aim of degrading the UE 1's \ac{SNR}, defined as $\mathrm{SNR}_1=|h_1|^2/\sigma^2$, where $\sigma^2$ is the receiver noise variance. Similarly, the UE $k$'s SNR is given as $\mathrm{SNR}_k=|h_k|^2/\sigma^2$.
The reason for minimizing the \ac{SNR} rather than the \ac{SINR} is to give the impression to the UE that its channel is blocked, which can happen naturally. 

\section{Destructive Beamforming with Perfect CSI}
This section is dedicated to the retrieval of the optimal \ac{SNR}-degrading $\ppsi$ under the assumption of perfect \ac{CSI} at the hacker.
The achieved \ac{SNR} degradation is intended as a benchmark to assess the \ac{RIS}'s \ac{SNR}-degradation capabilities and the impact of \ac{CEE}.
The \ac{SNR}-degrading $\ppsi$ acts as a destructive beamforming vector and will be obtained in two scenarios: a single-UE scenario, where the \ac{RIS} degrades \ac{UE}~$1$'s \ac{SNR} disregarding the possible presence of other \acp{UE}, and a multiple-UE scenario, where additional constraints are imposed to guarantee a minimum \ac{SNR} level to the other UEs. In the single-UE scenario, the most \ac{SNR}-degrading $\ppsi$ is found by solving the following optimization problem:

\begin{subequations}\label{single user original perfect}
\begin{align}
 \minimize_{\hat{\ppsi}}\, &~\hat{\ppsi}^\hermit \RR_1 \hat{\ppsi}\\
  \text{subject~to} &~ \left|\hat{\psi}_n\right|= 1, ~ n=1,\ldots, N , \label{single user original perfect:b}\\
  &~ \hat{\psi}_{N+1}=1, \label{last element 1}
\end{align}
\end{subequations}
where 
\begin{align}
    \hat{\ppsi}&=\left[\ppsi^\top , 1\right]^\top, \quad
      \RR_1 = \begin{bmatrix}
      \Breve{\hh}_1\Breve{\hh}^{\hermit}_1 & \Breve{\hh}_1 h_{\text{s},1}\\
     h_{\text{s},1}^{*} \Breve{\hh}^{\hermit}_1 &  \left|h_{\text{s},1}\right|^2 \label{eq:RR1}
      \end{bmatrix}.
\end{align}
This problem is non-convex due to the unit modulus constraints in \eqref{single user original perfect:b}. 
Rather than employing a \ac{SDR}-based approach, we apply a penalty-based \ac{CCP} approach \cite{pan2018robust} due to its lower computational complexity.
First, unit modulus constraints can be decoupled into a chain of inequalities such as $1 \leq |\hat{\psi}_n|^2 \leq 1,~n=1,\ldots, N$.
Then, the leftmost inequality is convexified by substituting $|\hat{\psi}_n|^2$ with its first-order Taylor expansion around the local point $\hat{\psi}_n^{(r)}$, which is the solution found at iteration $r$.
Finally, we define a penalty variable $\mathbf{d} = [d_1,\ldots,d_{2N}]^\top$ and introduce it into the previously defined inequalities.
Finally, the constraint \eqref{single user original perfect:b} can be redefined as 
\begin{subequations}
  \begin{align}
    &\left|\hat{\psi}_n\right|^2 \leq 1 + d_n~,~n=1,\ldots, N \label{convex constr a }\\
    &\left|\hat{\psi}_n^{(r)}\right|^2 - 2\Re\left( \hat{\psi}_n^*\hat{\psi}_n^{(r)}\right) \leq d_{N+n} -1 ~,~n=1,\ldots, N \label{convex constr b }.
\end{align}  
\end{subequations}
Problem \eqref{single user original perfect} can then be recast into the convex problem
\begin{subequations}\label{single user relaxed perfect}
\begin{align}
 \text{P}1 : ~ \minimize_{\hat{\ppsi},\dd }\,\,\, &~\hat{\ppsi}^\hermit \RR_1 \hat{\ppsi}+\lambda^{(r)} ||\mathbf{d}||\\
     \text{subject to}~& \eqref{convex constr a }, \eqref{convex constr b },\\
     &\hat{\psi}_{N+1}=1, \\
     &d_n \geq 0, \quad n=1,\ldots,2N, \label{eq:d-0}
\end{align}
\end{subequations}
where $\lambda^{(r)}$ is a multiplying factor, adjusting the impact of the penalty term  onto the objective function, at iteration $r$.

The multiple-UE optimization problem follows the same rationale but has $K-1$ additional minimum-\ac{SNR} constraints:
\begin{subequations}\label{multiple user  perfect}
\begin{align}
\minimize_{\hat{\ppsi} }\, &~\hat{\ppsi}^\hermit \RR_1 \hat{\ppsi}\\
  \text{subject to} &~ \hat{\ppsi}^\hermit \RR_k \hat{\ppsi} \geq \gamma_k\sigma^2, ~ k=2,\ldots,K, \label{minimum SNR og}\\
  &\left|\hat{\psi}_n\right|= 1, ~ n=1,\ldots, N , \label{single user original perfect:b2}\\
     &~ \hat{\psi}_{N+1}=1, \label{last element 1}
\end{align}
\end{subequations}
where $\RR_k$ is defined as in \eqref{eq:RR1} but with a different index and $\gamma_k$ is the minimum \ac{SNR} requirement for UE~$k$.
This problem is non-convex due to the presence of the constraints in \eqref{minimum SNR og}.
However, it can be convexified by substituting $\hat{\ppsi}^\hermit \RR_k \hat{\ppsi}$ with its first-order Taylor expansion around $\hat{\ppsi}^{(r)}$ :  
\begin{align}\label{min SNR taylor}
  \hat{\ppsi}^{(r)\hermit}\RR_k \hat{\ppsi}^{(r)} - 
  2\Re \left( \ppsi^{(r)\hermit} \RR_k \hat{\ppsi}\right) 
  \leq t_k-\gamma_k\sigma^2 ,
\end{align}
where $t_k$ is the penalty term for this constraint.
Problem \eqref{multiple user  perfect} can then be rewritten as 
\begin{subequations}\label{multiple user  perfect convex}
\begin{align}
\text{P}2:~\minimize_{\hat{\ppsi},\dd,\mathbf{t}} &~\hat{\ppsi}^\hermit \RR_1 \hat{\ppsi}+\lambda^{(r)} ||\mathbf{d}||+ \omega^{(r)}||\mathbf{t}||\\
  \text{subject to} &~ \eqref{min SNR taylor}, ~ k=2,\ldots,K, \\
  &\eqref{convex constr a },~\eqref{convex constr b }, ~\eqref{eq:d-0}\\
     & \hat{\psi}_{N+1}=1, \label{last element 1}\\
     &t_k \geq 0 , \quad k=2,\ldots,K,
\end{align}
\end{subequations}
where $\mathbf{t} = [t_2,\ldots,t_K]^\top$ and
$\omega^{(r)}$ is the multiplying factor associated with the penalty term $||\mathbf{t}||$. Choosing a fixed value of $\gamma_k$ would undermine the fairness of the performance comparison across different values of $N$: a certain minimum SNR is easy to ensure when $N$ is large, but it would be hard or impossible to guarantee said SNR when $N$ is low. We then define $\gamma_k$ as a fixed percentage of the maximum SNR achievable by each UE: its value is retrieved by solving the following problem 
\begin{subequations}\label{gamma definition}
    \begin{align}
    \gamma_k = c~ \maximize_{ \hat{\PPsi}}\, & ~\Tr\left(\RR_k\hat{\PPsi}\right)\\
     \text{subject to} ~& \hat{\PPsi}_{n,n}=1 ,~n=1,\ldots, N+1,\\
     &\hat{\PPsi}\succeq 0,~\text{rank}\left(\hat{\PPsi}\right) =1,
\end{align}
\end{subequations}
where $\hat{\PPsi}=\hat{\ppsi}\hat{\ppsi}^\hermit$ and $0<c\leq 1$. By relaxing the rank-one constraint, this problem becomes convex and can be directly solved by general-purpose convex optimization solvers.

We now assume that the RIS has a constant-magnitude channel (e.g., line-of-sight (LoS))  to both the AP and  \ac{UE} $1$: The reflected path can then be denoted as
$\Breve{\hh}_1 = \rho_\text{r}
        [e^{j\angle \Breve{h}_1(1)},\ldots,e^{j\angle \Breve{h}_1(N)}]^\top$,
 where %
 $\rho_\text{r}$ is the channel magnitude.
 The cascaded channel in \eqref{cascaded channel2} can now be written as 
 \begin{align} \label{eq:cascadedLOS}
  h_{\text{s},1} + \Breve{\hh}_1^\hermit \ppsi = \left| h_{\text{s},1}\right| e^{j\angle  h_{\text{s},1}} + \rho_\text{r}\sum_{n=1}^N e^{-j \angle \Breve{h}_1(n) } e^{ \angle \psi_n}.
 \end{align}
Under these assumptions, problem \eqref{single user original perfect} can be solved in closed form, as stated in the following lemma.
\begin{customlemma}{1}
Let the RIS choose $\xi\in[0,2\pi)$ and let the $n$-th \ac{RIS} phase-shift be $\angle \psi_n = \angle   h_{\text{\normalfont s},1} + \angle  \Breve{h}_1(n) + \left(n-\frac{N+1}{2}\right)\xi + \pi$.
 Then, the cascaded channel magnitude becomes  
 \begin{align}
     \left|  h_{\text{\normalfont s},1} +  \Breve{\hh}_1^\hermit \ppsi \right| =\left| \left|  h_{\text{\normalfont s},1}\right| - \rho_\text{\normalfont r}\frac{\sin(N \xi/2)}{\sin(\xi/2)}\right|.
 \end{align}
 If $|h_{\text{\normalfont s},1}|\geq N\rho_\text{\normalfont r}$, then  problem \eqref{single user original perfect} outcome is $\mathrm{SNR}_1=\left(\left|  h_{\text{\normalfont s},1}\right|-N\rho_\text{\normalfont r}\right)^2/\sigma^2$, obtained for $\xi=0$. If $\left| h_{\text{\normalfont s},1}\right| \leq N \rho_\text{\normalfont r}$, $\mathrm{SNR}_1$ can be nullified for some $\xi$.
\end{customlemma}
\textbf{Proof:}
The summation $\sum_{n=1}^N e^{-j \angle \Breve{h}_1(n) } e^{ \angle \psi_n}$ in \eqref{eq:cascadedLOS} can be rewritten as $\rho_\text{r}\sin(N \xi/2)/\sin(\xi/2)$, thanks to the geometric series formula and Euler's formula. The latter has peaks around $\xi=0$ with a maximum equal to $N\rho_\text{r}$.
If $|h_{\text{s},1}|$ is higher than this peak value, the best the \ac{RIS} can do is to choose $\xi=0$: on the other hand, if $|h_{\text{s},1}|$ is lower than the peak, the RIS can choose a $\xi$ such that SNR$_1=0$. \hfill\rule{1.5ex}{1.5ex}
 
\section{Imperfect CSI: Robust Optimization}\label{imperfect CSI theory}
We will now relax the perfect \ac{CSI} assumption by introducing \ac{CEE} onto the static path. This choice is motivated by the fact that this path is the hardest one to estimate from the hacker's point of view.
We can denote the cascaded channel in \eqref{cascaded channel2} as $h_k= \hat{h}_{\text{s},k} + \Delta h_{\text{s},k} + \Breve{\hh}^{\hermit}_k\ppsi$,
where $\Delta h_{\text{s},k}$ denotes the \ac{CEE}.
We adopt a bounded error model \cite{gao2021robust}, where the error can be anything satisfying
$|\Delta h_{\text{s},k}| \leq \epsilon_{\text{s},k}, k=1,\ldots, K$, and the upper limit $\epsilon_{\text{s},k}$ is assumed to be known by the hacker.
Under these assumptions, problem \eqref{single user original perfect} becomes
\begin{subequations}\label{single robust original}
\begin{align}
     \minimize_{ \ppsi} \,\, &\max_{~|\Delta h_{\text{s},1}|\leq\epsilon_{\text{s},1}}  ~\left|\hat{h}_{\text{s},1} +\Delta h_{\text{s},1} + \Breve{\hh}^{\hermit}_1\ppsi\right|^2 \\
     \text{subject to} ~& |\psi_n| = 1, ~ n=1,\ldots, N. 
\end{align}
\end{subequations}
As usual, when dealing with min-max problems, we define an auxiliary variable $a$ and recast the problem in epigraph form:
\begin{subequations}
    \begin{align}
        \minimize_{ \ppsi,a}\,\, & ~ a \\
     \text{subject to} ~& \left|\hat{h}_{\text{s},1} +\Delta h_{\text{s},1} + \Breve{\hh}^{\hermit}_1\ppsi\right|^2 \leq a, \ \forall~|\Delta h_{\text{s},1}|\leq \epsilon_{\text{s},1}, \label{geq schur} \\
     &|\psi_n| = 1, ~ n=1,\ldots, N, \\
     &a \geq 0.
    \end{align}
\end{subequations}
The presence of \ac{CSI} uncertainties makes \eqref{geq schur} a constraint of infinite cardinality, as $\Delta h_{\text{s},1}$ can take an infinite number of values.
To address this issue, we use the Schur complement \cite{DiRenzo2020robust} to rewrite the constraint in \eqref{geq schur} as
\begin{align}
\begin{bmatrix}
    a & x\\
    x^* & 1
\end{bmatrix} \succeq 0, \quad \forall~|\Delta h_{\text{s},1}|\leq\epsilon_{\text{s},1} ,
\end{align}
where $x= \hat{h}_{\text{s},1} +\Delta h_{\text{s},1} + \Breve{\hh}^{\hermit}_1\ppsi$.
We then apply Nemirovski's lemma to equivalently express the constraint as \cite{eldar2004robust}
\begin{align}\label{Nemirovsky constraint}
    \begin{bmatrix}
        a-\xi & \hat{x}   \\
        \hat{x}^* & 1  
    \end{bmatrix} \succeq 0 ,
    ~
    \begin{bmatrix}
        \xi  & \epsilon_{\text{s},1} \\
        \epsilon_{\text{s},1}& 1 \\
    \end{bmatrix} \succeq 0 ,
\end{align}
where $\hat{x} = \hat{h}_{\text{s},1} + \Breve{\hh}^{\hermit}_1\ppsi$ and $\xi\geq 0$ is an auxiliary variable.
Finally, we define the robust version of problem P$1$ as
\begin{subequations}
    \begin{align}
       \text{P}1' : ~ \minimize_{ \ppsi,a,\xi,\dd}\,\, & ~ a + \lambda^{(r)} ||\mathbf{d}|| \\
     \text{subject to} ~& \eqref{convex constr a },~\eqref{convex constr b },~\eqref{eq:d-0},~\eqref{Nemirovsky constraint}.
    \end{align}
\end{subequations}
The multi-user scenario optimization problem can be defined by adding $K-1$ minimum \ac{SNR} constraints to P$1'$.\footnote{To obtain a fair comparison, this robust approach uses the same minimum \ac{SNR} $\gamma_k$ as its perfect CSI counterpart.} We then obtain the problem
\begin{subequations}\label{multiple robust original}
\begin{align}
     \minimize_{ \ppsi,a,\xi,\dd}\, & ~ a + \lambda^{(r)} ||\mathbf{d}|| \\
     \text{subject to} ~& \eqref{convex constr a },~\eqref{convex constr b },~\eqref{eq:d-0},~\eqref{Nemirovsky constraint}, \\
    &\left|\hat{h}_{\text{s},k} +\Delta h_{\text{s},k} + \Breve{\hh}^{\hermit}_k\ppsi\right|^2 \geq \gamma_k \sigma^2, \label{QoS nonconvex}\\
      & \forall~|\Delta h_{\text{s},k}|\leq\epsilon_{\text{s},k},~k=2, \ldots, K \nonumber .
\end{align}
\end{subequations}
The constraint \eqref{QoS nonconvex} is once again non-convex and of infinite cardinality. Hence, we first approximate $|h_k|^2$ with a local lower bound.

\begin{customlemma}{2}
 Let $\ppsi^{(r)}$ be a feasible point for problem \eqref{multiple robust original} at iteration $r$. In that case,  $|h_k|^2$ can be lower bounded as
\begin{align}\label{lemma bound}
    |h_k|^2 \geq h_{\text{\normalfont s},k}^{*}h_{\text{\normalfont s},k} + h_{\text{\normalfont s},k}^{*} \Breve{\hh}^{\hermit}_k\ppsi + \ppsi^\hermit\Breve{\hh}_k h_{\text{\normalfont s},k} +c_k, 
\end{align}
where $c_k= \ppsi^{(r)\hermit} \Breve{\hh}_k\Breve{\hh}^{\hermit}_k\ppsi  +  \ppsi^\hermit \Breve{\hh}_k\Breve{\hh}^{\hermit}_k\ppsi^{(r)} - \ppsi^{(r)\hermit} \Breve{\hh}_k\Breve{\hh}^{\hermit}_k\ppsi^{(r)}$.  
\end{customlemma}
\textbf{Proof:} Any complex scalar variable $\delta$ can be lower bounded as \cite{pan2018robust} 
    $|\delta|^2 \geq \delta^{(r)*}\delta + \delta^*\delta^{(r)} - \delta^{(r)*}\delta^{(r)}$
for any $\delta^{(r)}$. By choosing $\delta = h_{\text{s},k}+\Breve{\hh}^{\hermit}_k\ppsi$ and $\delta^{(r)} = h_{\text{s},k}+\Breve{\hh}^{\hermit}_k\ppsi^{(r)}$ we obtain \eqref{lemma bound}. \hfill\rule{1.5ex}{1.5ex}

Given that $h_{\text{s},k}= \hat{h}_{\text{s},k} + \Delta h_{\text{s},k}$, \eqref{QoS nonconvex} can be reformulated as 
\begin{align}\label{Delta S procedure}
    &\Delta h_{\text{s},k}^{*} \Delta h_{\text{s},k} + 2\Re \left (\left( \Breve{\hh}^{\hermit}_k \ppsi  +\hat{h}_{\text{s},k}\right)^*\Delta h_{\text{s},k}
    \right) \nonumber\\   
    &+f_k \geq \gamma_k\sigma^2, \forall~|\Delta h_{\text{s},k}| \leq \epsilon_{\text{s},k},
\end{align}
where $f_k =\hat{h}_{\text{s},k}^{*} \hat{h}_{\text{s},k}  + \hat{h}_{\text{s},k}^{*} \Breve{\hh}^{\hermit}_k\ppsi+ \ppsi^\hermit\Breve{\hh}_k\hat{h}_{\text{s},k} +c_k$.
We are now able to get rid of $\Delta h_{\text{s},k}$ by applying the S-procedure \cite{luo2004multivariate}. If we define the auxiliary variable $\alpha$, \eqref{Delta S procedure} can be equivalently expressed as 
\begin{align}\label{S-procedure constraint}
    &\begin{bmatrix}
        1 + \alpha & \Breve{\hh}^{\hermit}_k \ppsi  +\hat{h}_{\text{s},k} \\
         \left( \Breve{\hh}^{\hermit}_k \ppsi  +\hat{h}_{\text{s},k}\right)^* & f_k-\gamma_k \sigma^2 - \alpha \epsilon_{\text{s},k}^2
    \end{bmatrix} \succeq 0.
\end{align}
Finally, we are able to rewrite problem \eqref{multiple robust original} as 
\begin{subequations}\label{multiple robust convex}
\begin{align}
    \text{P}2' : ~ \minimize_{ \ppsi,a,\xi,\alpha,\dd} \, & ~ a + \lambda^{(r)} ||\mathbf{d}|| \\
     \text{subject to} ~& \eqref{convex constr a },~\eqref{convex constr b },~\eqref{eq:d-0},~ \eqref{Nemirovsky constraint},~\eqref{S-procedure constraint},\\
     &\alpha \geq 0 .
\end{align}
\end{subequations}
All the problems presented in the last two sections are solved by Algorithm $1$, where  $\mathcal{P}=\{ \text{P}1 , \text{P}2 , \text{P}1',\text{P}2'\}$.

\begin{algorithm} [t!]
  \caption{CCP based algorithm for \ac{RIS} destructive beamforming optimization}
  \begin{algorithmic}[1]
    \STATE \textbf{Initialize:} $\ppsi^{(0)},\omega^{(0)},\lambda^{(0)}$, set $r=0$
    \IF{P$_p$ $\neq$ P2}
       \STATE$\mathbf{t}=\mathbf{0}$
    \ENDIF
    \REPEAT
        \STATE Compute $\ppsi^{(r+1)}$ by solving P$_p \in \mathcal{P}$, $r \leftarrow r+1$
         \STATE $\lambda^{(r)}=\min\left(\mu\lambda^{(r-1)},\lambda_\text{max} \right)$, 
         \IF{P$_p$=P2}
            \STATE$\omega^{(r)}=\min\left(\mu\omega^{(r-1)},\omega_\text{max}\right)$
         \ENDIF
    \UNTIL{$||\dd|| \leq \nu $, $\left\Vert\ppsi^{(r)}-\ppsi^{(r-1)}\right\Vert \leq  \nu$,$||\mathbf{t}||\leq \nu$}
    \STATE \textbf{Output:} $\ppsi_\text{opt}$
  \end{algorithmic}
\end{algorithm}

\section{Numerical Results}
We now demonstrate the destructive beamforming that a malicious \ac{RIS} can perform in different situations.
The Monte-Carlo simulations are obtained using $200$ independent channel realizations.
The transmitter is located at $(10,0)$\,m whereas the \ac{RIS} center is located at $(50,100)$\,m. There are $K=2$ UEs located at $(300,0)$\,m and $(300,50)$\,m.
The static paths $\hh_{\text{s},k}$ are modeled as Ricean fading, with a K-factor equal to $10$, $\HH_t$ is assumed to be a LoS channel whereas $\hh_{\text{r},k}$ is Rayleigh fading, The large-scale shadowing coefficient is $\beta_k = -30 - 20 \log_{10}(d_k/d_0)$, where $d_k$ is the distance between the transmitter, either the BS or the RIS, and UE $k$ and $d_0=1$\,m.
The noise power at all the UEs is $-70$\,dB.
The performances of our algorithm are compared against a \ac{MRT} benchmark, where the RIS phase-shifts are defined as $\psi_n = \Breve{h}_2(n)/|\Breve{h}_2(n)|, n=1,\ldots, N$. This approach leverages the large channel dimensionality (i.e., favorable propagation) to divert the signal away from UE $1$.
The \ac{CEE} bounds are defined as $\epsilon_{\text{s},k} = \eta |h_{\text{s},k}|, k=1,\ldots,K$, where $0 \leq \eta \leq 1$.
The parameters , $\lambda_{\max}$ and $\omega_{\max}$are equal to $10^4$, $\mu=1.5$, and $\nu=10^{-3}$

\begin{figure}[t!]
\centering
\resizebox{0.45\textwidth}{!}{
%
%
\definecolor{mycolor1}{rgb}{0.00000,0.44700,0.74100}%
\definecolor{mycolor2}{rgb}{0.85000,0.32500,0.09800}%
\definecolor{mycolor3}{rgb}{0.9290 ,0.6940 ,0.1250}%
\definecolor{mycolor4}{rgb}{0.4940, 0.1840, 0.5560}%
\definecolor{mycolor5}{rgb}{0.4660, 0.6740, 0.1880}%
\begin{tikzpicture}

\begin{axis}[%
width=3in,
height=1.8in,
at={(0.758in,0.582in)},
scale only axis,
xmin=0,
xmax=30,
xlabel style={font=\color{white!15!black}},
xlabel={$N$},
ymin=-61,
ymax=20,
ylabel style={font=\color{white!15!black}},
ylabel={SNR $1$ [dB]},
axis background/.style={fill=white},
axis x line*=bottom,
axis y line*=left,
xmajorgrids,
ymajorgrids,
legend style={at={(1,0.7)},legend cell align=left, align=left, draw=white!15!black}
]
\addplot [color=mycolor4, line width=1.0pt,mark=square,mark options={solid}]
  table[row sep=crcr]{%
0   11.33123\\
2	12.5\\
5	13.23123\\
10	14.13123\\
15	14.733123\\
20  15.33123\\     
25  16.13123\\  
30  17.33123\\  
};
\addlegendentry{random}
\addplot [color=mycolor5, line width=1.0pt,dashed,mark=triangle,mark options={solid}]
  table[row sep=crcr]{%
0   11.53123\\
2	12.5\\
5	13.023123\\
10	13.533123\\
15	14.0733123\\
20  14.533123\\     
25  15.13123\\  
30  15.833123\\  
};
\addlegendentry{MRT}

\addplot [color=mycolor3, line width=1.0pt]
  table[row sep=crcr]{%
0   11.33123\\
2	11.33123\\
5	11.33123\\
10	11.33123\\
15	11.33123\\
20  11.33123\\     
25  11.33123\\  
30  11.33123\\  
};
\addlegendentry{No RIS}

\addplot [color=mycolor2, line width=1.0pt,mark=*,mark options={solid}]
  table[row sep=crcr]{%
0   11.33123\\
2	9.95\\
5	8.8\\
10	7.1\\
15	4.315\\
20  2.49\\     
25  0.022476\\  
30  -3.42187\\  
};
\addlegendentry{P2,$c=0.9$}
\addplot [color=mycolor1, line width=1.0pt,,mark=triangle,mark options={solid}]
  table[row sep=crcr]{%
0   11.33123\\
2	5.95\\
5	-4.2\\
10	-18.8\\
15	-37\\
20  -47.67\\     
25  -52.0467\\  
30  -55.8\\  
};
\addlegendentry{P1}

\end{axis}
\end{tikzpicture}%
}
\vspace{-2mm}
     \caption{SNR degradation analysis with perfect CSI.}
     \label{perfect CSI performances}
     \vspace{-5mm}
 \end{figure}
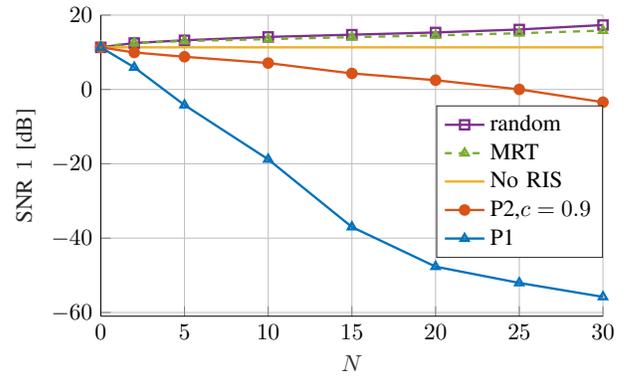
\begin{figure}[t!]
 \centering
\resizebox{0.45\textwidth}{!}{
%
%
\definecolor{mycolor1}{rgb}{0.00000,0.44700,0.74100}%
\definecolor{mycolor2}{rgb}{0.85000,0.32500,0.09800}%
\definecolor{mycolor3}{rgb}{0.9290 ,0.6940 ,0.1250}%
\definecolor{mycolor4}{rgb}{0.4940, 0.1840, 0.5560}%
\definecolor{mycolor5}{rgb}{0.4660, 0.6740, 0.1880}%
\begin{tikzpicture}

\begin{axis}[%
width=0.8\linewidth,
height=4cm,
at={(0.58in,2.54in)},
scale only axis,
xmin=0,
xmax=30,
xlabel style={font=\color{white!15!black}},
xlabel={$N$},
ymin=-60,
ymax=20,
ylabel style={font=\color{white!15!black}},
ylabel={SNR $1$ [dB]},
axis background/.style={fill=white},
axis x line*=bottom,
axis y line*=left,
legend columns=2,
xmajorgrids,
ymajorgrids,
legend style={at={(1,0.8)},legend cell align=left, align=left, draw=white!15!black}
]

\addplot [color=mycolor4, line width=1.0pt,mark=square,mark options={solid}]
  table[row sep=crcr]{%
0   11.33123\\
2	12.5\\
5	13.23123\\
10	14.13123\\
15	14.733123\\
20  15.33123\\     
25  16.13123\\  
30  17.33123\\  
};
\addlegendentry{random}
\addplot [color=mycolor5, line width=1.0pt,dashed,mark=triangle,mark options={solid}]
  table[row sep=crcr]{%
0   11.53123\\
2	12.5\\
5	13.023123\\
10	13.533123\\
15	14.0733123\\
20  14.533123\\     
25  15.13123\\  
30  15.833123\\  
};
\addlegendentry{MRT}

\addplot [color=mycolor3, line width=1.0pt]
  table[row sep=crcr]{%
0   11.33123\\
2	11.33123\\
5	11.33123\\
10	11.33123\\
15	11.33123\\
20  11.33123\\     
25  11.33123\\  
30  11.33123\\  
};
\addlegendentry{No RIS}
\addplot [color=mycolor1, line width=1.0pt,dashed,mark=square,mark options={solid}]
  table[row sep=crcr]{%
0   11.33123\\
2	6.7\\
5	-1.9\\
10	-13\\
15	-31.4\\
20  -41.2\\     
25  -45\\  
30  -50.5\\  
};
\addlegendentry{$\eta = 1$}
\addplot [color=mycolor1, line width=1.0pt,dashed,mark=triangle,mark options={solid}]
  table[row sep=crcr]{%
0   11.33123\\
2	6.5\\
5	-2.85\\
10	-17\\
15	-33\\
20  -42.8\\     
25  -46.54\\  
30  -53\\  
};
\addlegendentry{$\eta = 0.1$}

\addplot [color=mycolor1, line width=1.0pt,,mark=*,mark options={solid}]
  table[row sep=crcr]{%
0   11.33123\\
2	5.95\\
5	-4.2\\
10	-19.8\\
15	-37\\
20  -50.67\\     
25  -54.0467\\  
30  -58.18\\  
};
\addlegendentry{P1}

\end{axis}
\node (bar2) at (5,5.3) {(a)};

\begin{axis}[%
width=0.8\linewidth,
height=4cm,
at={{(0.58in,0.4in)}},
scale only axis,
xmin=0,
xmax=30,
xlabel style={font=\color{white!15!black}},
xlabel={$N$},
ymin=-5,
ymax=18,
ylabel style={font=\color{white!15!black}},
ylabel={SNR $1$ [dB]},
axis background/.style={fill=white},
axis x line*=bottom,
axis y line*=left,
legend columns=2,
xmajorgrids,
ymajorgrids,
legend style={at={(0.62,0.36)},legend cell align=left, align=left, draw=white!15!black}
]

\addplot [color=mycolor4, line width=1.0pt,mark=square,mark options={solid}]
  table[row sep=crcr]{%
0   11.33123\\
2	12.5\\
5	13.23123\\
10	14.13123\\
15	14.733123\\
20  15.33123\\     
25  16.13123\\  
30  17.33123\\  
};
\addlegendentry{random}
\addplot [color=mycolor5, line width=1.0pt,dashed,mark=triangle,mark options={solid}]
  table[row sep=crcr]{%
0   11.53123\\
2	12.5\\
5	13.023123\\
10	13.533123\\
15	14.0733123\\
20  14.533123\\     
25  15.13123\\  
30  15.833123\\  
};
\addlegendentry{MRT}

\addplot [color=mycolor3, line width=1.0pt]
  table[row sep=crcr]{%
0   11.33123\\
2	11.33123\\
5	11.33123\\
10	11.33123\\
15	11.33123\\
20  11.33123\\     
25  11.33123\\  
30  11.33123\\  
};
\addlegendentry{No RIS}
\addplot [color=mycolor2, line width=1.0pt,dashed,mark=square,mark options={solid}]
  table[row sep=crcr]{%
0   11.33123\\
2	11.5\\
5	10.9\\
10  9.7\\
15	7.2\\
20  5.6\\     
25  3.85\\  
30  0.8\\  
};
\addlegendentry{$\eta=0.2$}
\addplot [color=mycolor2, line width=1.0pt,dashed,mark=triangle,mark options={solid}]
  table[row sep=crcr]{%
0   11.33123\\
2	10.5\\
5	9.78\\
10	8.2\\
15	5.36\\
20  3.2\\     
25  1.26\\  
30  -2\\  
};
\addlegendentry{$\eta=0.1$}
\addplot [color=mycolor2, line width=1.0pt,mark=*,mark options={solid}]
  table[row sep=crcr]{%
0   11.33123\\
2	9.95\\
5	8.8\\
10	7.1\\
15	4.315\\
20  2.49\\     
25  0.022476\\  
30  -3.42187\\  
};
\addlegendentry{P2,$c=0.9$}

\end{axis}
\node (bar3) at (5,-0.1) {(b)};

\end{tikzpicture}%
}
\vspace{-2mm}
     \caption{SNR degradation analysis under imperfect CSI. }
    \label{imperfect CSI}
    \vspace{-5mm}
 \end{figure}
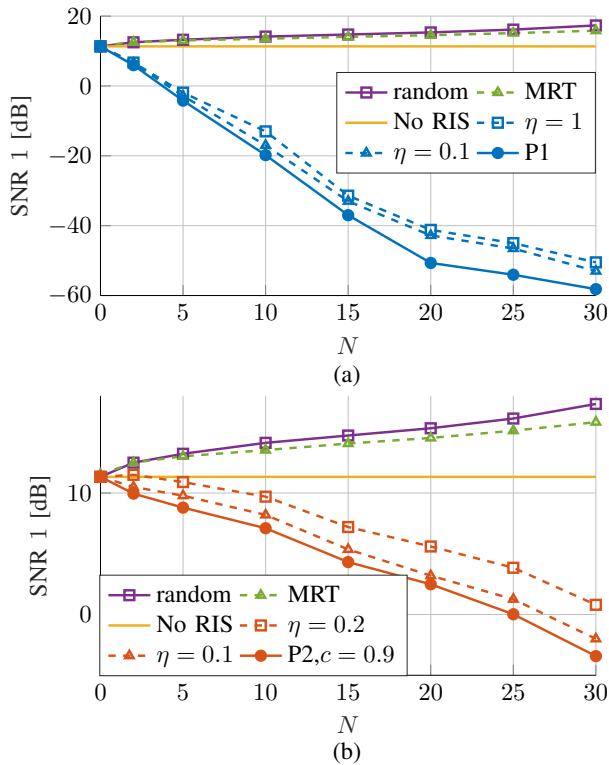

\subsection{Perfect \ac{CSI}}

The impact of the malicious \ac{RIS} is directly dependent on the static path strength relative to $\Breve{\hh}_k$. As $N$ grows, the magnitude of the reflected path grows as well and the malicious RIS' silent attack can become more powerful.
This trend is showcased in Fig. \ref{perfect CSI performances}, where $\mathrm{SNR}_1$ is shown as a function of the number of RIS elements, $N$.
We compare our algorithm with two low complexity approaches, namely the previously mentioned \ac{MRT} and a random approach, where each phase-shift $\psi_n$ takes a random value in $[0,2\pi)$.
The first feature that we can observe is that our algorithm drastically outperforms the other approaches, first and foremost the random approach, since we know that a user SNR in the presence of random RIS phase-shifts grows linearly with $N$ \cite{bjornson2024introduction}.
Secondly, the \ac{MRT} approach, albeit being slightly better than the random one, still does not give a SNR reduction, as we see that said SNR is still always above the ``no RIS'' curve, equal to $|h_{\text{s},1}|^2 /\sigma^2$.
As for our proposed algorithm, the presence of a minimum SNR requirement has a sizable impact on the overall performance. If the presence of other UE is disregarded, we see that $5$ reflective elements are enough to obtain a $20$\,dB SNR reduction, whereas if the presence of other UE is taken into account, the RIS needs at least $30$ elements to replicate the same SNR reduction.


\subsection{Imperfect \ac{CSI}}
We will now investigate the impact of channel uncertainties on the \ac{RIS}' \ac{SNR}-degrading action, using the results presented in the previous section as a benchmark.
Fig.~\ref{imperfect CSI} shows the effect of \acp{CEE} as a function of $N$ for different values of $\epsilon_{\text{s},k}$.
The presence of \acp{CEE} does not alter the inverse proportionality between SNR $1$   and $N$, however, a larger \ac{CEE} magnitude obviously leads to worse performance.
In Fig.~\ref{imperfect CSI}, we can observe how the destructive beamforming performance in the single-user case is robust against CEEs.  \acp{CEE} have a non-negligible impact, even when $\eta=1$ only $15$ reflective elements are sufficient to obtain a $40$\,dB reduction.
This can be ascribed to the fact that in P$1$ and P$1'$, the most important thing is to phase-align the destructive paths rather than knowing the static one.
The same cannot be said about the multi-user cases. Indeed, Fig. \ref{imperfect CSI}(b) shows that even relatively small errors can severely hinder the RIS's destructive capability.
This is partially explained by the very stringent minimum SNR introduced.

\section{Conclusions}
We investigated a novel scenario where a hacker takes control of a \ac{RIS} with the intention of degrading the SNR of a specific UE while preserving the SNR of all other UEs.
Assuming perfect \ac{CSI}, the \ac{RIS} phase-shift vector design strategy is presented, with a closed-form solution for LoS channels. 
We then devised a robust optimization approach.
Our simulations demonstrate that our algorithms can dramatically reduce a UE SNR and how minimum SNR constraints limit this action, especially in the presence of CEEs.

\bibliographystyle{IEEEtran}
\bibliography{IEEEabrv,auxiliary/biblio}
\end{document}